%% file: main.tex
\documentclass[10pt,twocolumn]{article}\usepackage[]{graphicx}\usepackage[]{xcolor}
\makeatletter
\def\maxwidth{ %
  \ifdim\Gin@nat@width>\linewidth
    \linewidth
  \else
    \Gin@nat@width
  \fi
}
\makeatother

\definecolor{fgcolor}{rgb}{0.345, 0.345, 0.345}

\usepackage{framed}
\makeatletter
 {\par\unskip\endMakeFramed%
 \at@end@of@kframe}
\makeatother

\definecolor{shadecolor}{rgb}{.97, .97, .97}
\definecolor{messagecolor}{rgb}{0, 0, 0}
\definecolor{warningcolor}{rgb}{1, 0, 1}
\definecolor{errorcolor}{rgb}{1, 0, 0}

\usepackage{alltt}
\usepackage[top=0.75in,bottom=0.75in,left=0.5in,right=0.5in]{geometry}
\usepackage[numbers,comma,sort&compress]{natbib}
\usepackage{times}
\usepackage{amsmath}
\usepackage{amssymb}
\usepackage{amsthm}
\usepackage{empheq}
\usepackage{mathtools}
\usepackage{graphicx}
\usepackage{tikz}
\usepackage{nameref}
\usepackage{textcomp}
\usepackage{gensymb}
\usepackage{siunitx}
\usepackage{nth}
\usepackage{bibunits}
\usepackage{url}
\usepackage[utf8]{inputenc}
\usepackage[OT1]{fontenc}

\usepackage{chngcntr}

\usepackage[switch]{lineno}

\makeatletter
\g@addto@macro{\UrlBreaks}{\UrlOrds}
\makeatother

\newcommand*\patchAmsMathEnvironmentForLineno[1]{%
   \expandafter\let\csname old#1\expandafter\endcsname\csname #1\endcsname
   \expandafter\let\csname oldend#1\expandafter\endcsname\csname end#1\endcsname
   \renewenvironment{#1}%
      {\linenomath\csname old#1\endcsname}%
      {\csname oldend#1\endcsname\endlinenomath}}%
\newcommand*\patchBothAmsMathEnvironmentsForLineno[1]{%
   \patchAmsMathEnvironmentForLineno{#1}%
   \patchAmsMathEnvironmentForLineno{#1*}}%
\AtBeginDocument{%
\patchBothAmsMathEnvironmentsForLineno{equation}%
\patchBothAmsMathEnvironmentsForLineno{align}%
\patchBothAmsMathEnvironmentsForLineno{flalign}%
\patchBothAmsMathEnvironmentsForLineno{alignat}%
\patchBothAmsMathEnvironmentsForLineno{gather}%
\patchBothAmsMathEnvironmentsForLineno{multline}%
\patchAmsMathEnvironmentForLineno{empheq}%
}

\usepackage[symbol*,perpage,hang,flushmargin]{footmisc}
\DefineFNsymbols*{fsym}[math]{\dagger\ddagger{\dagger\dagger}{\ddagger\ddagger}}
\setfnsymbol{fsym}

\graphicspath{{figures/}{diagrams/}}

\makeatletter
\def\input@path{{./}{figures/}}
\makeatother

\usepackage{caption}
\usepackage[labelformat=simple]{subfig}

\usepackage{floatrow}
\newdimen\figrasterwd
\figrasterwd\textwidth

\makeatletter
\renewcommand{\maketitle}{\bgroup\setlength{\parindent}{0pt}
	\begin{flushleft}
		\@title
		\\[1em]
		\@author
	\end{flushleft}\egroup
}
\makeatother

\renewenvironment{abstract}
{\par\noindent\textbf{\abstractname}\\ \ignorespaces}
{\par\medskip}

\makeatletter
\renewcommand\@biblabel[1]{#1.}
\makeatother

\usepackage[immediate]{silence}
\usepackage{sectsty}
\allsectionsfont{\sffamily}
\DeactivateWarningFilters[temp] 
\IfFileExists{upquote.sty}{\usepackage{upquote}}{}
\begin{document}

\include{definitions}


\title{\vspace{-3em} \LARGE \sffamily \textbf{%
    Historical constraints on the evolution of efficient color naming
}}
\author{\small
        Colin R. Twomey\textsuperscript{\tiny{1,2}}*,
        David H. Brainard\textsuperscript{\tiny{3}},
	\& Joshua B. Plotkin\textsuperscript{\tiny{2}}
	\\[0.75em]
	{\scriptsize \textsuperscript{\tiny{1}} Data Driven Discovery Initiative, University of Pennsylvania, Philadelphia, PA, USA} \\
	{\scriptsize \textsuperscript{\tiny{2}} Department of Biology, University of Pennsylvania, Philadelphia, PA, USA} \\
	{\scriptsize \textsuperscript{\tiny{3}} Department of Psychology, University of Pennsylvania, Philadelphia, PA, USA} \\
	{\scriptsize * Corresponding author. Email: crtwomey@sas.upenn.edu (C.R.T.)}\\
}

\twocolumn[
\maketitle
\vspace{0.5em}
\renewcommand{\abstractname}{\vspace{-\baselineskip}}
\begin{abstract}
\textbf{%
\noindent Color naming in natural languages is not arbitrary: it reflects efficient partitions of perceptual color space \cite{regier2007} modulated by the relative needs to communicate about different colors \cite{twomey2021}.
These psychophysical and communicative constraints help explain why languages around the world have remarkably similar, but not identical, mappings of colors to color terms.
Languages converge on a small set of efficient representations.
But languages also evolve \cite{greenhill2010}, and the number of terms in a color vocabulary may change over time.
Here we show that history, i.e.\ the existence of an antecedent color vocabulary, acts as a non-adaptive constraint that biases the choice of efficient solution as a language transitions from a vocabulary of size $n$ to $n + 1$ terms.
Moreover, as vocabularies evolve to include more terms they explore a smaller fraction of all possible efficient vocabularies compared to equally-sized vocabularies constructed {\em de novo}.
This path dependence on the cultural evolution of color naming presents an opportunity.
Historical constraints can be used to reconstruct ancestral color vocabularies, allowing us to answer long-standing questions about the evolutionary sequences of color words, and enabling us to draw inferences from phylogenetic patterns of language change.
}
\end{abstract}
\vspace{1.5em}
]

\section*{Introduction}
Are our mental representations of the world anything like those of our ancestors?
How could we ever know?
On the one hand this is a question about cognition, and it can be addressed by comparative study of extant non-human primates \cite{corballis2017}, at least over long evolutionary timescales.
But for modern humans, over thousands rather than millions of years \cite{powell2009b,greenhill2010}, this is a question about culture and how cultures evolve over time.
Color naming -- a language community's mapping of colors to color terms -- has become a model system for studying the link between cognition and culture, thanks in part to extensive field work \cite[e.g.][]{brown1954,kay2009,maclaury1997,heider1972,levinson2000,lindsey2015,gibson2017}
and theory \cite{jameson1997,yendrik2001,regier2007,jager2007,zaslavsky2018,twomey2021} spanning decades.
Far from being arbitrary, color naming follows common patterns in linguistic communities around the world \cite{berlin1969,kay2009} that reflect near-optimal partitions of perceptual color space \cite{regier2007} modulated by cross-cultural differences in the needs to communicate about different colors \cite{twomey2021}.
Color naming systems can therefore be seen as efficient allocations of terms to colors \cite{zaslavsky2018,twomey2021}.

Although recent work has explained extant patterns in color naming based on cognitive constraints and communicative needs \cite{regier2007,gibson2017,twomey2021}, this leaves open the question of how color naming in a linguistic community is expected to change over time.
Early work by Berlin \& Kay posited that color naming systems will follow a stereotyped evolutionary sequence in which color terms appear successively in a (nearly) fixed order \cite{berlin1969}.
This hypothesis was subsequently revised based on a larger collection of empirical color naming data, broadening the number of potential evolutionary pathways that color vocabularies may take as they increase in size \cite{kay1999,kay2009}.
The overall conceptual model -- that of a limited number of evolutionary transitions from one stage to the next -- has remained largely the same.
But one key issue has received comparatively little attention: the potential for path-dependent effects, i.e.\ the possibility that the next stage of evolutionary development may depend on the path taken in the prior stage.

Are path-dependent evolutionary trajectories for color vocabularies consistent with the theory of efficient color naming?
Here we study this question by analyzing how history has constrained color naming within an established theoretical framework that predicts color vocabularies as efficient solutions to a representation problem.
Whereas the geometry of perceptual color space and communicative demands on colors determine the landscape of efficient color naming systems in this framework, there might nonetheless be path dependence -- i.e.\ the current choice of an efficient solution may constrain the accessibilities of potential future choices.
One possibility is that, depending on where in color space a new color term is introduced, any one of several equally efficient vocabularies may be reached.
This would be inconsistent with the B\&K conceptual model, which posits that some successor vocabularies are unreachable (or at least unlikely) given some prior vocabularies, and it would suggest that color term evolution is more flexible than previously thought.
Alternatively, an extant color vocabulary may constrain where a newly introduced color term can become established, thus biasing which efficient vocabularies can be reached from an initial vocabulary.
In this case, the initial vocabulary would act as a non-adaptive\footnote{Note that non-adaptive does not imply maladaptive. In evolutionary theory, a maladaptive trait is harmful to the fitness of its bearer, while a non-adaptive trait has neither a direct cost nor a benefit.} constraint on cultural evolution, producing path-dependent effects in color naming systems.

\section*{Results}

\subsection*{Color categories}

\begin{figure*}[htb]
\centering
\includegraphics[width=\textwidth]{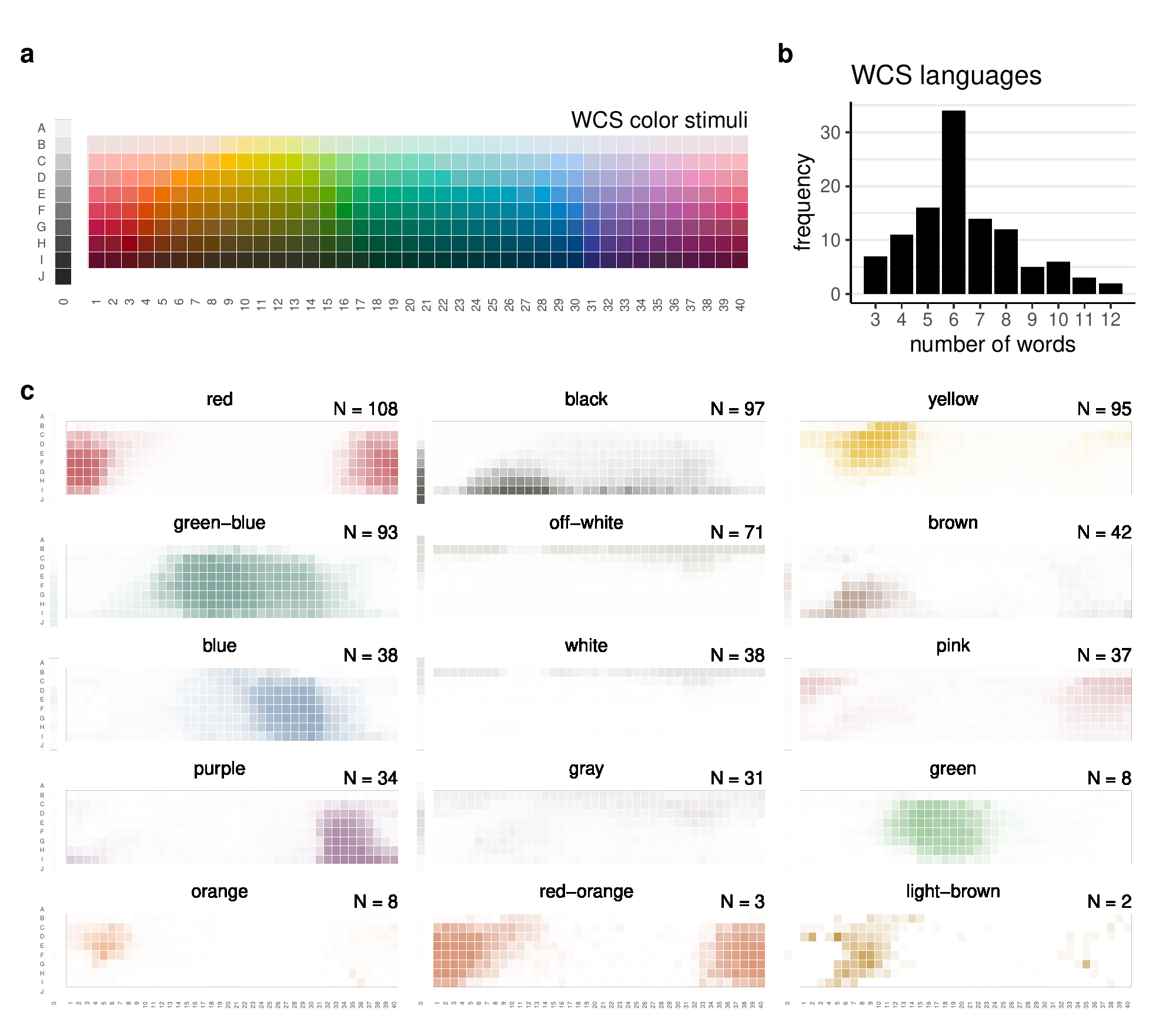}
\caption{%
\textbf{%
	A cross-language dictionary of 15 color categories (``color words'') derived from clustering all color terms in the empirical WCS languages.
}
(\textbf{a}) The 330 color stimuli shown to native language speakers participating in the WCS study \cite{berlin1969,kay2009}.
(\textbf{b}) Distribution of vocabulary sizes across the WCS languages. Vocabularies of size $n \leq 8$ are well-represented ($N \geq 7$ languages); vocabularies of size $9$, $11$, and $12$ words are rare ($N = 5$, $3$, and $2$ languages, respectively).
(\textbf{c}) The mapping of WCS color stimuli from \textbf{a} to each color category.
An English-language description of the average color for that category is given as the title of each panel, and each mapping is displayed with the corresponding color.
Specifically, each panel shows the average (across languages) conditional probability, $p(x|\hat{x})$, that a native speaker's use of the term for color category $\hat{x}$ in their language refers to WCS color stimulus $x$.
The conditional probability for each stimulus is shown on a zero to one scale, from transparent to the average WCS color of the category.
The number, $N$, of WCS languages containing a term corresponding to each color category is shown in the upper right of each panel.
}\label{fig:word-clusters}
\end{figure*}

To study the evolution of color naming systems we must first identify color categories that apply across languages.
We take a quantitative approach similar to Lindsey \& Brown \cite{lindsey2009}, using the World Color Survey (WCS) color naming experiments to identify cross-language clusters of color terms as universal color categories (Fig.~\ref{fig:word-clusters}; see Methods:~\nameref{methods:wcs}).
In the WCS study, roughly 25 speakers from each of 110 languages were asked to name 330 standardized color stimuli.
Averaging across speakers, this provides a \textit{color term map} for each language -- namely, the chance that a speaker will use each term to describe a given color stimulus.
Using Bayes' rule we also compute the associated \textit{color stimulus map} for each term in a language -- namely, the probability that each stimulus will be named by that term.

We define cross-language \textit{color categories} as clusters of color terms identified via modularity maximization \cite{newman2004} (Methods:~\nameref{methods:dictionary}).
We apply a modularity maximization algorithm to the pairwise Earth Mover's distance (EMD) between the stimulus maps for all terms in all languages (Methods:~\nameref{methods:distance}).
The EMD, or 1\textsuperscript{st} Wasserstein distance, measures the minimum cost of transforming one term's distribution over colors into another term's distribution.
Modularity maximization identified 15 term clusters, which we call color categories, across the 110 languages in the empirical WCS.
These clusters include the 11 color categories previously identified by Berlin \& Kay, as well as additional categories for orange (occurring in 8 languages), red-orange (3 languages), and light-brown (2 languages), along with a distinction between ``off-white'' (71 languages) and ``white'' (38 languages).
A dictionary of which WCS color stimuli (Fig.~\ref{fig:word-clusters}a) are typically mapped to each of the 15 color categories is shown in Fig.~\ref{fig:word-clusters}c).

Using this dictionary, we assigned each term in each language to one of the identified universal color categories by minimizing the EMD between the category-average stimulus map and the language's stimulus map.
If multiple terms in the same language are assigned to the same color category, we consider those terms to be synonyms and treat them as a single term.
The resulting non-synonymous terms we call the language's color words, and each word is identified with a unique (cross-language) color category.
We refer to the complete set of color words in a language as its color vocabulary, and the number of such words is its vocabulary size (Fig.~\ref{fig:word-clusters}b).

\subsection*{Evolutionary transitions}

What happens when a new term is added to an existing vocabulary of size $n$?
To study this question we use an established model of color naming that captures the tendency of color vocabularies to evolve towards efficient partitions of color space \cite{regier2007} modulated by language-specific communicative needs for colors \cite{gibson2017,twomey2021}.

Let $X$ be a random variable that represents observable colors in the WCS experiment, taking on values in the Fig.~\ref{fig:word-clusters}a array of 330 WCS color stimuli.
Associated with each language's color terms, denoted $\widehat{X}$, is a probabilistic mapping of colors to terms, $p(\hat{x}|x)$, that gives the probability that a speaker will use term $\hat{x} \in \widehat{X}$ to refer to color $x \in X$.
While any mapping of colors to color terms is possible, only some mappings are efficient.
An efficient mapping minimizes the average distortion introduced by using a color term, where distortion, $\dbreg{\vec{x}}{\vec{\hat{x}}}$, measures the perceptual dissimilarity between a color $x$ and the color typically associated with term $\hat{x}$.
Both $x$ and $\hat{x}$ are associated with coordinates in a perceptually uniform color space, $\vec{x}$ and $\vec{\hat{x}}$, for measuring distances.
The average distortion introduced by a mapping of colors to terms also depends on the language-specific need to communicate each color, $p(x)$, and it is calculated as
\begin{align}
    D\left[ p(\hat{x}|x) \right] &= \sum_{x,\hat{x}} p(\hat{x}|x) \, p(x) \, \dbreg{\vec{x}}{\vec{\hat{x}}}.
\end{align}
Zero average distortion can be achieved using a vocabulary with a unique term matched to every color, i.e.\ where $|\widehat{X}| = |X|$ and, for every $x$, there exists an $\hat{x}$ such that $\vec{\hat{x}} = \vec{x}$.
But the representational cost -- i.e.\ a rate corresponding to the minimum number of bits needed to encode $\widehat{X}$ on average, per observation of $X$ -- increases with increasing vocabulary size and specificity.
This cost is quantified by the mutual information between $X$ and $\widehat{X}$:
\begin{align}
    R\left[p(\hat{x}|x)\right] = \I{X}{\widehat{X}}
    = \sum_{x,\hat{x}} p(\hat{x}|x) \, p(x) \log \frac{p(\hat{x}|x)}{p(\hat{x})}.
\end{align}
The total cost of a choice of mapping, $p(\hat{x}|x)$, is given by $R + \beta D$, where $\beta \in [0,\infty)$ parameterizes the trade-off between representational cost and average distortion (in bits per unit of distortion).
As $\beta$ increases, the specificity of the mapping increases and the average distortion decreases.
In this way, $\beta$ and the number of terms, $|\widehat{X}|$, co-determine the representational cost of any given mapping from colors to terms (see SI~Fig.~\ref{si-fig:rd-example}).
A rate-distortion efficient vocabulary, for a particular choice of $\beta$ and $|\widehat{X}|$, is a choice of mapping $p(\hat{x}|x)$ that minimizes $R[p(\hat{x}|x)] + \beta D[p(\hat{x}|x)]$.

For fixed communicative needs, $p(x)$, and rate-distortion trade-off, $\beta$, the following dynamics
\begin{align}
	p_{t+1}(\hat{x}) &= \sum_x p_t(\hat{x}|x) \, p(x),\label{eq:vocab-dynm-1}\\
	p_{t+1}(\hat{x}|x) &\propto p_t(\hat{x}) e^{-\beta \dbreg{\vec{x}}{\vec{\hat{x}}}},\label{eq:vocab-dynm-2}\\
	\vec{\hat{x}}_{t+1} &= \sum_x \vec{x} \, p_t(x|\hat{x})\label{eq:vocab-dynm-3},
\end{align}%
provably converge to locally-optimal rate-distortion efficient vocabularies as $t \rightarrow \infty$ \cite{banerjee2005}.
Rates below this efficient frontier for any given choice of allowable distortion (parameterized by $\beta$) are impossible.
Illustrated diagrammatically,
\begin{center}
\begin{tikzpicture}

\fill[gray!40] (0,2) .. controls (1,0) and (2,0) .. (3,0) -- (0,0);
\draw[thick,->] (0,0) -- (0,2.5);
\draw[thick,->] (0,0) -- (3.5,0);
\draw (0,2) -- (-0.1,2) node[anchor=east] {\scriptsize $H(X)$};
\node at (-0.5,1) {$R$};
\node at (1.5,-0.4) {$D$};
\node at (2.78,1.5) {\small $\circ$ $p_0(\hat{x}|x)$};
\draw[dashed,->] (2.2,1.35) .. controls (2.2,1.3) and (2,0.7) .. (1.5,0.61) -- (1.17,0.6);
\draw (0.4,1.1) -- (1.6,0.075);
\node[circle] at (0.99,0.59) {\small $\bullet$};
\node at (1.2,0.9) {\small $\beta$};
\node at (2.45,0.65) {\small $t \rightarrow \infty$};

\end{tikzpicture}
\end{center}
these dynamics move an initial choice of color term mapping, $p_0(\hat{x}|x)$, from the feasible region (shown in white) towards the efficient frontier -- the boundary between feasible (white) and infeasible (gray) solutions, for any choice of trade-off between $R$ and $D$, parameterized by $\beta$.
(See SI~Sec.~\ref{si-sec:rd-example} for an example that also varies $|\widehat{X}|$).

By systematically introducing a new term at low initial frequency, varying its initial focal position over the set of 330 WCS stimuli, and running forward the vocabulary dynamics (Eqs.~\ref{eq:vocab-dynm-1}~--~\ref{eq:vocab-dynm-3}), we determined the number of unique ($n + 1$)-word vocabularies that a given $n$-word vocabulary may generate, as well as the relative size of the basin of attraction for each (the number of focal positions that converge to the same $(n + 1)$-word vocabulary; see Methods:~\nameref{methods:successor}).
By assigning the terms of the resulting $(n + 1)$-word vocabulary to the closest matching color category, we can identify when the newly introduced term results in a set of non-synonymous $(n + 1)$ color words, as well as any changes in meaning between the $n$- and $(n + 1$)-word vocabularies.

\begin{figure*}[htb]
\centering
\includegraphics[width=\textwidth]{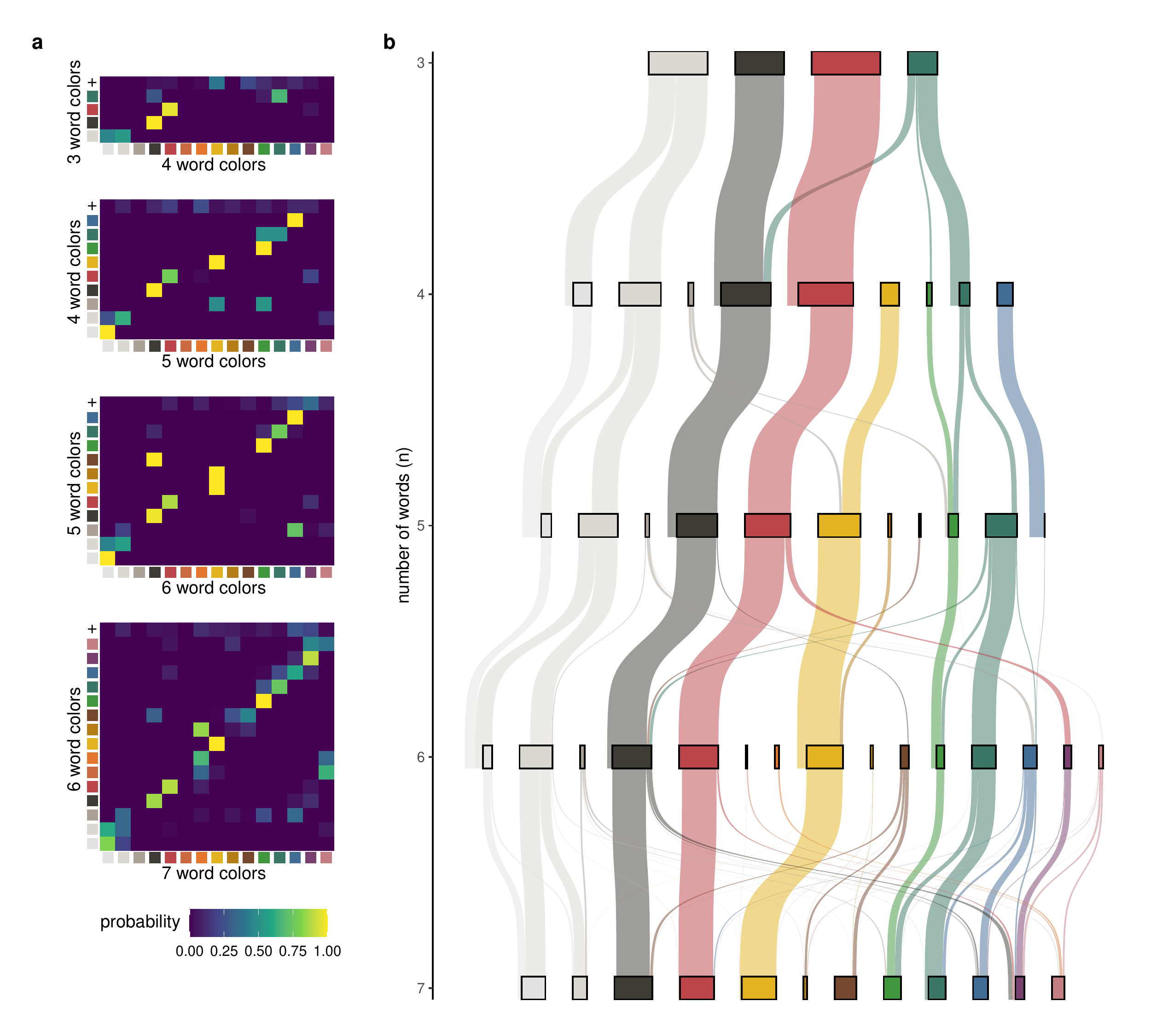}
\caption{%
\textbf{%
Transitions from $n$- to $(n + 1)$-word vocabularies, based on historically constrained empirical data.
}
(\textbf{a}) Semantic shifts in the meaning of color terms.
Each row shows the probabilities that a term in a $n$-word color vocabulary, originally identified with one color word from the dictionary (Fig.~\ref{fig:word-clusters}), is identified with each of the color words (columns) after the addition of a new term.
For example, a term corresponding to the color category ``off-white'' in a $3$-word vocabulary may become associated with either ``off-white'' or ``white'' in a resulting $4$-word vocabulary.
The rows marked $+$ indicate the identification probabilities (columns) of the added term.
(\textbf{b}) Sankey diagram of semantic shifts in color word meaning after the introduction of a new term.
Each row shows the relative frequency of color words in WCS vocabularies of a given vocabulary size (y-axis), where box size correspond to the frequency of words among languages and box colors correspond to Fig.~\ref{fig:word-clusters}c.
Semantic mappings between stages are shown by connecting lines, where the thickness of each line corresponds to the fraction of times a transition occurs under the modeled rate-distortion dynamics.
}\label{fig:transitions}
\end{figure*}

Using the WCS languages as initial vocabularies we systematically probed the introduction of new terms, and we identified the probability that a given word will transition in meaning as the vocabulary size increases (Fig.~\ref{fig:transitions}).
When a new word is introduced, it may displace or carve out a new ``niche'' in color space that impacts the mappings of existing color words.
For example, a term that was identified as green-blue in a language of vocabulary size $n=5$ may shift in meaning and become identified with the color category ``green'' after a newly introduced term establishes as ``blue.''
Or, the existing green-blue term could become ``blue,'' while the newly introduced term becomes ``green.''
We quantified the probability of a change in meaning as the frequency with which a given color word in extant $n$-word vocabularies (corresponding to rows in Fig.~\ref{fig:transitions}a) is identified with a given color word in the expected $(n + 1)$-word vocabulary (columns in Fig.~\ref{fig:transitions}b) after a new term is introduced (marginalizing over both languages and initial placements of the new term).

Color words vary in their susceptibility to semantic shift under addition of a new term.
Fig.~\ref{fig:transitions}b illustrates the transition probabilities shown in Fig.~\ref{fig:transitions}a diagrammatically: the thickness of the curves linking color words 
show the expected fraction of each type of semantic transition.
Red, black, and (once it appears) yellow are relatively stable in meaning over successive additions of new terms, whereas e.g.\ green-blue and blue are more prone to semantic change.
The broader off-white color category feeds into, and then is largely replaced by, a sharper white category, in a successional sequence of vocabularies from $n=3$ words to $n=7$ words.

\subsection*{Historical constraints on vocabularies}

\begin{figure*}[htb]
\centering
\includegraphics[width=\textwidth]{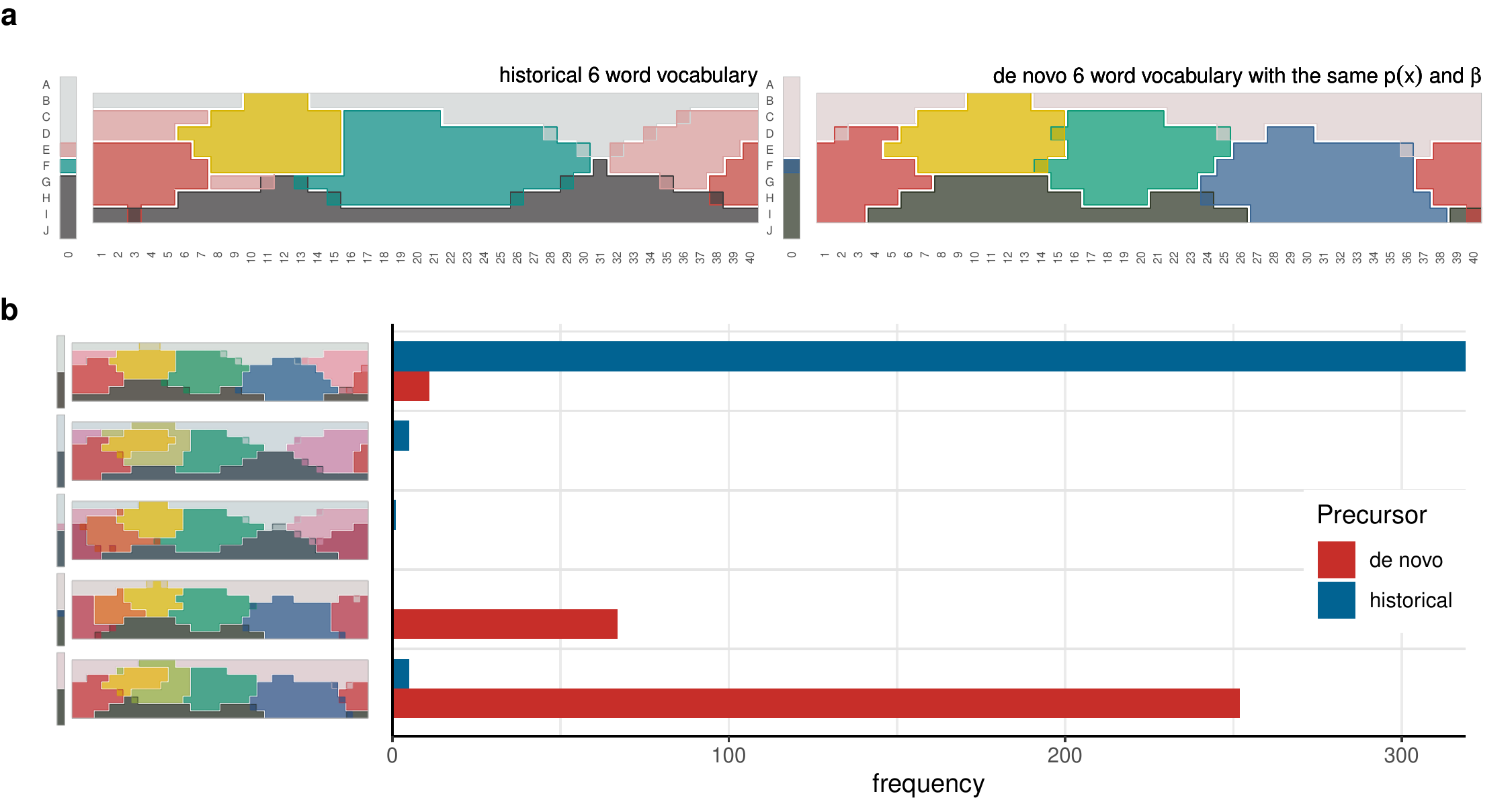}
\caption{%
\textbf{%
Historical constraints influence the likelihood of alternative efficient vocabularies.
}
(\textbf{a}) The rate-distortion efficient vocabulary corresponding to an empirically measured (``historical'') WCS color vocabulary ({\em left}) can differ from a rate-distortion efficient vocabulary constructed {\em de novo} using the same language-specific communicative needs, $p(x)$, and same need for precision, $\beta$.
In this example, one vocabulary contains a term for pink, while the other contains no term for pink and distinct terms for green and for blue.
(\textbf{b}) Adding a new term either to the {\em de novo} 6-word vocabulary (red bars) or to the historical 6-word vocabulary (blue bars) produces one of five different efficient $7$-word vocabularies (rows) under the rate distortion model.
The frequencies of these efficient solutions vary widely, and they depend strongly on the precursor language (red bars versus blue bars).
In particular, the historical precursor vocabulary with a single word for green-blue is most likely to develop separate words for green and blue (blue bars); whereas the {\em de novo} precursor vocabulary lacking pink is unlikely to introduce pink and more likely to introduce either light-green or orange (red bars).
}\label{fig:comparison}
\end{figure*}

While both $n$- and $(n + 1)$-word vocabularies are rate-distortion efficient, they are often not unique.
Which particular $(n + 1)$-word vocabulary will appear after addition of a new color term depends on the precursor $n$-word vocabulary.
For two efficient vocabularies constrained by the same communicative needs and rate-distortion tradeoff (Fig.~\ref{fig:comparison}a), the probabilities of likely successor vocabularies can vary widely (mean within-language SD in probability of successor vocabularies was 0.40~$\pm$~0.005~SE; Fig.~\ref{fig:comparison}b).
For example, an extant 6-word WCS vocabulary with words for white, black, yellow, pink, and green-blue, will most likely split green-blue into a term for green and a term for blue; whereas a similar efficient vocabulary that already has blue and green rather than green-blue, but no pink, is unlikely to next introduce pink but more likely to introduce light-green or orange (Fig.~\ref{fig:comparison}b).

\begin{figure*}[htb]
\centering
\includegraphics[width=\textwidth]{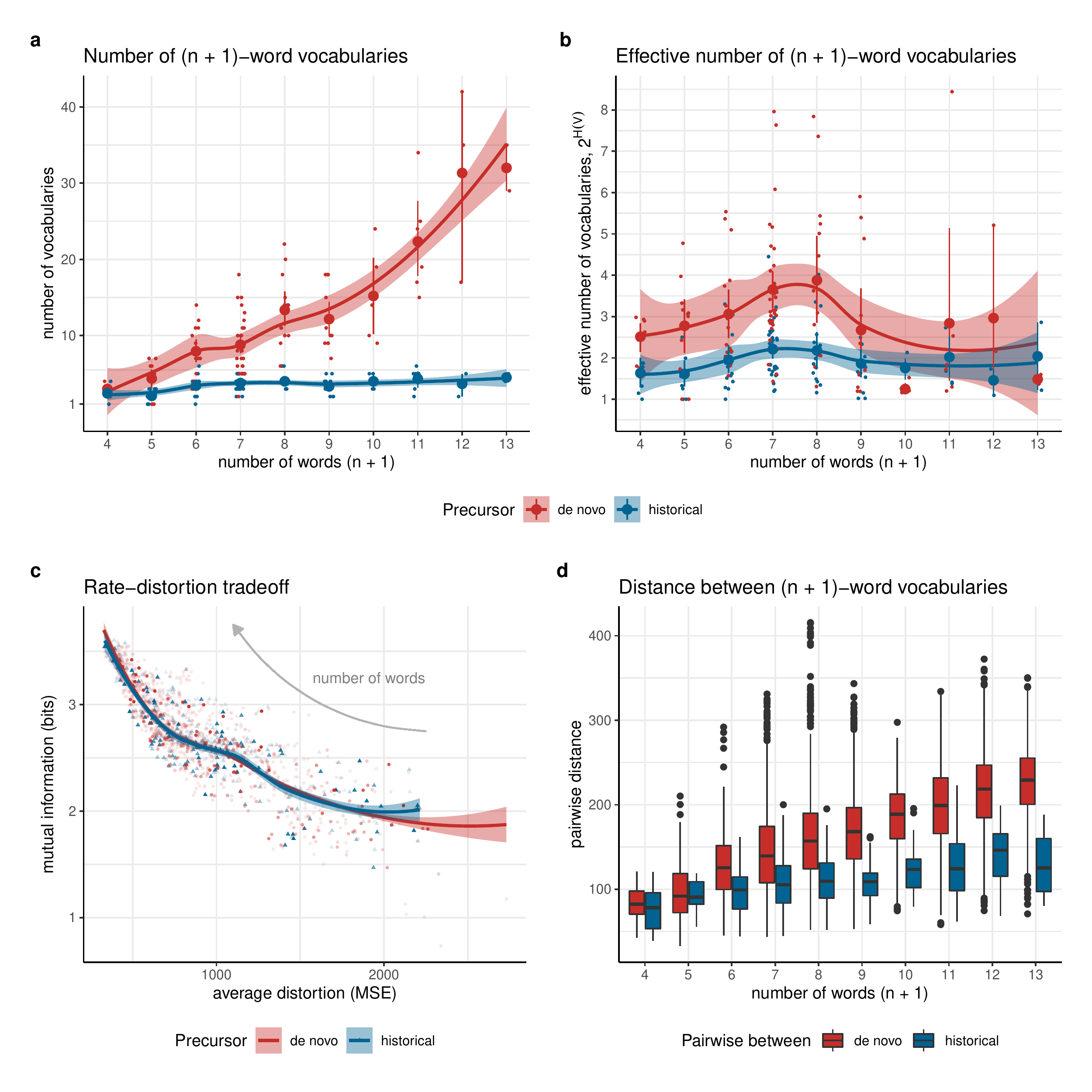}
\caption{%
\textbf{%
The number of possible efficient vocabularies exceeds the number of historically-constrained successor vocabularies.
}
(\textbf{a}) As vocabulary size (number of words) increases, the total number of possible efficient vocabularies (constructed {\em de novo}) increases while the number of historically constrained successor vocabularies remains relatively constant.
(\textbf{b}) The effective number of potential successor vocabularies (accounting for the relative frequency of each solution) as a function of vocabulary size.
Here, the ``effective number'' is computed as $2^{\h{V}}$, where $\h{V}$ is the entropy of the frequency distribution over the set $V$ of distinct successor vocabularies.
This quantity is largest and equal to the cardinality of $V$ when every successor vocabulary is equally likely.
(\textbf{c}) Historically constrained and {\em de novo} efficient vocabularies show no systematic difference in the quality of rate-distortion efficient solutions, as measured by their rate (mutual information) and average distortion. Lines show weighted (by solution frequency) LOESS regressions of rate as a function of distortion.
(\textbf{d}) Pairwise Wasserstein distance between historically constrained $(n + 1)$-word vocabularies and {\em de novo} vocabularies of size $n + 1$.
For small vocabulary sizes, the solution space explored is of similar size; as vocabulary size increases, {\em de novo} solutions span a larger space.
}\label{fig:modes-summary}
\end{figure*}

The strength of historical constraint on successor vocabularies increases as vocabulary size grows.
To measure this difference, we used the same communicative needs and rate-distortion tradeoff inferred for each WCS language, and constructed {\em de novo} all unique locally-optimal efficient vocabularies following the vocabulary dynamics in Eqs.~\ref{eq:vocab-dynm-1}~--~\ref{eq:vocab-dynm-3}.
This superset of each extant WCS language vocabulary contained one or more choices of efficient vocabulary possible given the inferred constraints, with the likelihood of each choice given by its basin of attraction (number of converged solutions based on $1,000$ random initial configurations).
The number of unique vocabularies generated by adding a term to an extant 3-word vocabulary is almost as large as the number of unique 4-word vocabularies generated {\em de novo}.
But the latter quickly outpaces the former as vocabulary size increases (Fig.~\ref{fig:modes-summary}a).
This remains true when weighting by the frequency of each potential successor vocabulary (Fig.~\ref{fig:modes-summary}b) -- at least up to vocabularies of size 8, after which the number of WCS languages is sparse (Fig.~\ref{fig:word-clusters}b) and uninformative -- and does not reflect any systematic difference in the quality of the rate-distortion efficient solutions achieved under {\em de novo} or historical constraints (Fig.~\ref{fig:modes-summary}c).
This also remains true when comparing the space spanned by the ensemble of historically constrained versus {\em de novo} successor vocabularies (Fig.~\ref{fig:modes-summary}d).
The space of all efficient vocabularies is much richer for large {\em de novo} vocabularies than for those realized in extant languages constrained by an evolutionary process of successive additions of terms.
To put this succinctly: history matters.

\subsection*{Ancestral vocabulary reconstruction}

Historical constraints on color vocabularies present an opportunity: they can be used to estimate likely ancestral languages.
Holding communicative needs and the community’s rate-distortion tradeoff constant, we can generate {\em de novo} $(n - 1)$-word vocabularies and compute the likelihood of their $[(n - 1) + 1]$-word successor vocabularies.
The likely ancestral vocabularies are those whose successor vocabularies are likely to produce the observed extant $n$-word vocabulary (Methods:~\nameref{methods:ancestral}).

\begin{figure*}[htb]
\centering
\includegraphics[width=\textwidth]{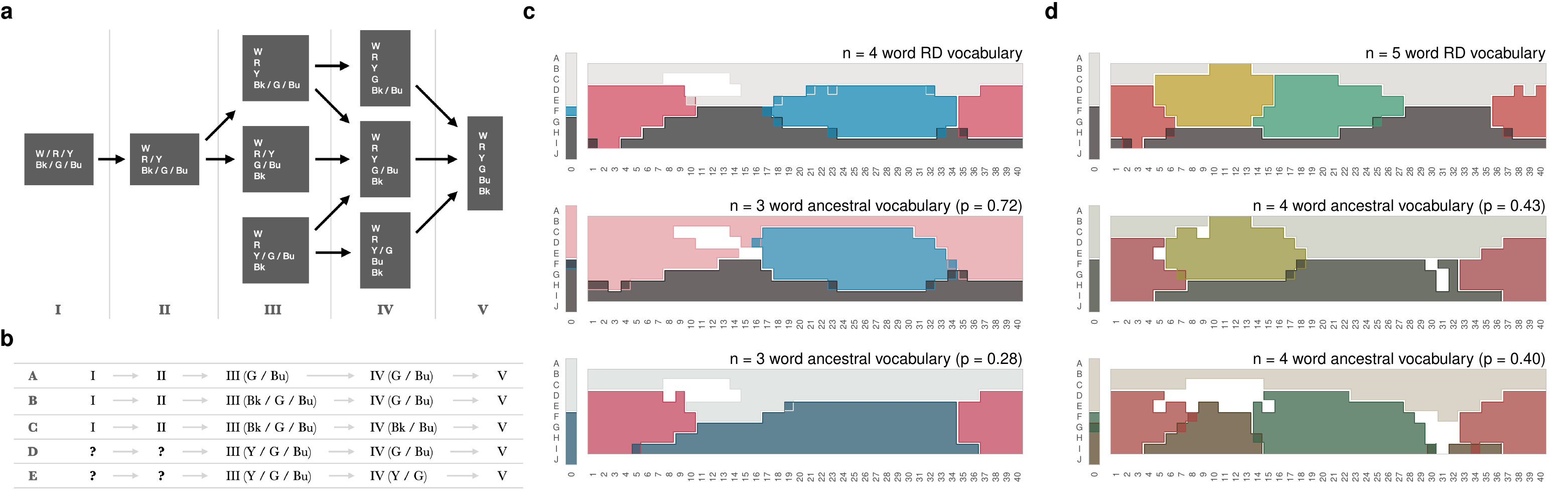}
\caption{%
\textbf{%
Ancestral language reconstructions for WCS languages, compared to hypothesized evolutionary transitions.
}
(\textbf{a}) Kay, Berlin, Maffi \& Merrifield (KBMM) \cite{kay1997,kay1999} classified the color vocabularies in the WCS data for languages containing between 2 and 6 terms, and hypothesized possible evolutionary transitions between them (arrows).
According to this classification, Stage I languages have two color words -- grouping white, red, and yellow (W/R/Y) as one term, and grouping black, green, blue (Bk/G/Bu) as a second term.
Stages I, II, and V vocabularies have one form each, while Stages III and IV WCS languages (4- and 5-word vocabularies, respectively) have three possible forms each.{\protect \footnote{Diagram adapted from \citet{kay1997} and \citet{kay1999}.\label{attribution}}}
(\textbf{b}) KBMM hypothesized several alternative evolutionary sequences, including the hypothesized most common ``mainline'' sequence, A; two additional sequences, B and C that vary at stage III; as well as two alternative sequences, D and E, that contain unknown stage-I and stage-II precursor vocabularies.\textsuperscript{\ref{attribution}}
(\textbf{c}) Likely rate-distortion efficient ancestral (precursor) languages for a stage III Y/G/Bu 4-word language ({\em top}).
Among the two possible reconstructions of ancestral 3-word languages, the more likely precursor ($p=0.72$, {\em middle}) does not correspond to any vocabulary observed in the WCS study; whereas the less likely precursor ($p=0.28$, {\em bottom}) corresponds to a stage-II language.
(\textbf{d}) For comparison, a more typical Stage IV G/Bu vocabulary ({\em top}) is shown along with two almost equally likely reconstructed ancestral languages that agree with the sequences hypothesized by KBMM.
The most likely precursor ($p=0.43$, {\em middle}) corresponds to a Stage III G/Bu vocabulary (sequence B), and the next most likely precursor ($p=0.40$, {\em bottom}) corresponds to a Stage III Bk/G/Bu vocabulary (sequence A).
}\label{fig:nminus1-vocabularies}
\end{figure*}

The Kay, Berlin, Maffi \& Merrifield (KBMM) conceptual model of color term evolution (Fig.~\ref{fig:nminus1-vocabularies}a) contains a long-standing mystery \cite{kay1999}: what are the potential precursor 3-word vocabularies for the observed 4-word vocabularies that contain words for white (W), red (R), black (Bk), yellow/green/blue (Y/G/Bu) (Fig.~\ref{fig:nminus1-vocabularies}b lines D and E)?
Using the ancestral reconstruction method described above, we computed all potential 3-word precursors that fully partition color space and are rate-distortion efficient, shown in Fig.~\ref{fig:nminus1-vocabularies}c.
The most likely such precursor has a combined term for red and white -- a vocabulary that is not observed among the languages surveyed in the WCS.
The only other efficient 3-term precursor is $\approx 39\%$ less likely, according to our analysis, 
although it appears among the languages sampled in the WCS: it has terms for white, red/yellow, and black/green/blue.
This example demonstrates the utility of directly reconstructing ancestral vocabularies: it can reveal ancestral states that may not be present in a sample of extant languages, or that may not persist to the present at all.

As a second example of this reconstruction method, we considered the possible precursors for a 5-word WCS vocabulary that falls along one of the ``mainline'' KBMM evolutionary sequences.
We identify three potential precursor vocabularies, with the two most likely shown in Fig.~\ref{fig:nminus1-vocabularies}d.
These correspond to a precursor with words for white, red/yellow, green/blue, and black (mainline path A), and a precursor with words for white, red, yellow, and black/green/blue (mainline path B).
These two precursors are almost equally likely, and each is $\approx 245\%$ more likely than the third alternative.
Thus the reconstructed vocabularies and their relative likelihoods allow us to distinguish among the potential evolutionary paths considered by KBMM, and can also reveal when ancestry is ambiguous.

\section*{Discussion}

Systems of color naming are constrained by perception and by communicative needs \cite{yendrik2001,jameson1997,regier2007,zaslavsky2018,gibson2017,twomey2021}.
Our study reveals a third, qualitatively different constraint on color vocabularies -- namely, history.
Languages evolve from precursor languages, and this alone produces quantifiable constraints on color vocabularies, introducing path dependence in color word evolution.
As new terms are added to an $n$-word vocabulary, the existing terms will shift in meaning and extent to accommodate.
We have shown that this flexibility is limited, so that not all efficient $(n + 1)$-word vocabularies are equally likely.
As color vocabularies increase in size, language communities explore only a small subset of the full range of efficient solutions that are possible.

These results establish that the theory of efficient color naming is consistent with a stereotypical set of evolutionary pathways, as hypothesized in early studies \cite{berlin1969,kay1978,kay1997,kay1999,kay2009}. 
Our analysis adds substantial detail to this hypothesis by quantifying the relative probabilities of different evolutionary pathways.
This perspective also provides a new answer to a long-standing question about ancestral precursors of particular languages in the WCS without invoking additional dynamics.
At the same time, our model is not incompatible with an alternative ``emergence hypothesis'' (EH) that posits that at early stages of development some color naming systems do not fully partition all of color space (see \citet{kay1999}).
Rather than assuming communicative needs are constant over time, as we have done in this work, efficient naming under EH suggests that some regions of color space had no communicative need whatsoever in the ancestral language.
Still, our results do suggest that color vocabularies of size 5 or larger are unlikely to emerge {\em de novo}.
Fig.~\ref{fig:modes-summary}a--d show that there is substantial overlap in historically constrained and {\em de novo} vocabularies only for those of size $n=4$ words.
Thus vocabularies larger than 4 words likely developed from an already efficient precursor, which is consistent with the few known examples of languages that cover only part of the color domain \cite[e.g.]{levinson2000,lindsey2015,lindsey2016}.

Many important questions remain.
For one, quantitative estimates of likely transitions in color vocabularies depend on where in color space new terms are introduced, which may not be uniform over the WCS color space as assumed in our analyses.
Are there biases -- cognitive, communicative, or historical -- on where new terms are likely to be introduced?

Second, while the present work investigates semantic shift when new terms are added to an existing vocabulary, for simplicity this analysis has been done while holding constant the communicative needs and $\beta$, the rate-distortion tradeoff parameter controlling the need for precision.
How might changes in the need for precision or the distribution of communicative needs across colors alter the course of cultural evolution?

Third, this study has focused on the addition of new terms to an existing language, but deletions might occur as well.
Recent work on color vocabulary phylogenies suggests that term loss may be more common than previously thought \cite{haynie2016}.
We observe one form of loss in our results: the addition of a term to a vocabulary can result in the loss of one color word in exchange for two new words.
A loss of this kind is due to semantic shift: e.g.\ a green-blue term becomes a green term after the newly added term establishes as blue.
But \citet{haynie2016} indicate a distinct form of loss: a deletion that strictly reduces the size of a color vocabulary, contra KBMM \cite{kay1997,kay1999}.
We have not treated this possibility in the present study.
If such losses are common, repeated cycles of term deletion and addition could allow languages to explore a larger fraction of the efficient vocabularies that are possible.
This process could still be investigated using the framework developed in this work, 
using a more general Markov process that allows for any number of $n$ to $n-1$ and $n-1$ to $n$ transitions, parameterized by the probability of a loss or gain at a vocabulary size of $n$.

Answering these questions will require better contact between dated language phylogenies and models of efficient color naming.
Color word presence/absence data of the kind used in \citet{haynie2016} are insufficient for this purpose: we require direct knowledge of the mappings between terms and colors in order to infer language-specific communicative needs for colors and the need for precision.
Whereas the world-spanning breadth of the WCS makes it ideal for answering questions about color naming similarities and differences across cultures, its depth in any particular language family is relatively shallow, limiting its utility for linking color naming to language phylogenies.
Future work could use the relatively high number of Austronesian languages within the WCS ($N=8$) in combination with recent dated phylogenies for the Austronesian language family \cite{gray2000} to study the questions of ancestral needs and term loss.
An alternative is the under-explored Mesoamerican Color Survey (MCS) collected by \citet{maclaury1997}, which catalogued 116 indigenous languages spoken in Mexico and Central America.
Our analysis of color term evolution gives further motivation to rehabilitate these data \cite{jameson2016,jameson2015} and develop dated language phylogenies for more Mesoamerican language families \cite[e.g.][for Uto-Aztecan]{greenhill2023}.
Our work also underscores the utility of longitudinal studies that sample color naming in a linguistic community over time \cite[e.g.][]{kuriki2017,zaslavsky2022}.

History likely constrains efficient representations in other aspects of language and culture aside from color -- such as kinship and spatial relations \cite{kemp2012,regier2015,kemp2018,gibson2019}, numeral systems \cite{xu2020}, and person systems \cite{zaslavsky2021}.
Our approach to quantifying historical constraints and inferring ancestral states could be used to study the evolution of these cultural systems as well.
There can be surprising utility in understanding how the elements of culture and language change over time; for example, using the colors reported in astronomical observations made thousands of years ago to estimate the fate of nearby stars \cite{neuhauser2022}.

Given the backdrop of evolutionary thought, our results are not entirely surprising.
The concept of phylogenetic constraint -- i.e.\ the idea that an organism's ancestry imposes limits on its evolutionary trajectory -- is fundamental to evolutionary biology.
\citet{gould1979} introduced the concept of ``spandrels,'' or non-adaptive evolutionary byproducts that arise as a consequence of selection for other traits, as an illustration of phylogenetic constraint.
Phylogenetic constraint can shape the pace and direction of evolution and limit the range of phenotypes that evolve within a lineage.
Quantifying these constraints has been essential for interpreting the patterns and processes of biological form and function.
Our work suggests that the same may be true of cultural evolution.
This supports the broader view that studying evolutionary processes can provide valuable insights into cultural systems (in color naming \cite[e.g.][]{komarova2007,xu2013} and in general \cite{dunn2011,dediu2013,levinson2014,creanza2017}), and vice versa.

\section*{Methods}

{\small

\subsection*{World Color Survey}\label{methods:wcs}

Using a standardized set of 330 color stimuli based on \citet{brown1954}, \citet{kay2009} catalogued color naming in 110 languages around the world.
Fieldworkers presented each of the color stimuli, one at a time in a fixed, randomized order, to on average 24 native speakers.
Ambient lighting was approximately controlled by presenting stimuli at noon in the shade.
The results of this survey are publicly available online from the WCS Data Archives (\path{https://www.icsi.berkeley.edu/wcs/data.html}).

\subsection*{Dictionary of color categories}\label{methods:dictionary}

Instead of clustering color terms across languages based on overlapping WCS speaker color maps (choices of words for a given color) as in \citet{lindsey2009}, we clustered based on inferred meanings of a given word, i.e.\ the probability a speaker is referring to a color given a choice of word.
We computed the probability, $p(x|\hat{x})$, that a speaker using a language's color term, $\hat{x}$, refers to a color, $x$, according to $p(x|\hat{x}) \propto p(x|\hat{x}) p(x)$, where $p(\hat{x}|x)$ is the WCS speaker-average color map and $p(x)$ is the inferred language-specific communicative needs from \citet{twomey2021}.
We then measured the dissimilarity, $D_{ij}$, as the EMD between every pair of color terms across all WCS languages.
Using the adjacency matrix $A_{ij} = \exp\left[-D_{ij}\right]$, we formulated the problem of identifying cross-language groupings of color terms (color term ``universals'') as a graph-theoretic ``community identification'' problem (a class of problems in network science concerned with identifying clusters, or ``communities,'' in a given network).
As in \citet{jackson2019} for identifying colexification patterns\footnote{Colexification refers to the use of a single word to represent multiple concepts in a language. Colexification patterns across languages can be used to estimate the degree of similarity between concepts \cite{francois2008,natale2021}.} across languages, we identified communities based on their modularity.
Unlike \citet{jackson2019}, the large size of $A$ prohibited the use of the exponential time linear integer programming formulation of the problem given by \citet{brandes2008}; instead, we used the well-known approximation method of \citet{clauset2004}, implemented in the R \texttt{igraph} package \cite{csardi2006}.

\subsection*{Earth Mover's Distance}\label{methods:distance}

Earth Mover's Distance (EMD) is a measure of the difference between two probability distributions, defined as the minimum cost of transforming one distribution into another, where cost is calculated as the sum of the distances between each point in the two distributions multiplied by the amount of mass moved.
In other words, the EMD measures the minimum amount of ``mass transportation'' required to move the points in one distribution to match the corresponding points in the other distribution.
We compute this quantity using the \texttt{emdist} R package based on the \citet{rubner1998} implementation.

\subsection*{Generating successor $n + 1$ vocabularies}\label{methods:successor}

Given a rate-distortion efficient vocabulary with $n$ words, candidate $n + 1$ term successor vocabularies were generated by systematically introducing (with low initial frequency equal to 1e-09) a new term with a focal color initialized to each WCS color stimulus (330 total) and running forward the rate-distortion dynamics given by Eqs.~\ref{eq:vocab-dynm-1}--\ref{eq:vocab-dynm-3} to equilibrium, while the communicative needs, $p(x)$, of the language, and need for precision, $\beta$, were held constant.
Candidate successor vocabularies generated in this way are not unique, and result in one or a small number of efficient solutions.
Non-unique (i.e.\ approximately identical) candidate successor vocabularies were judged based on the average squared distance (in CIE Lab space) between the focal colors of corresponding color terms between the two vocabularies, where correspondence was determined based on the minimum distance assignment problem matching focal colors from one vocabulary to the other.\footnote{The minimum distance assignment problem is a mathematical problem that involves finding the optimal assignment of objects from one set to another based on a given cost function. In this case, the cost function is the squared distance between the focal colors of corresponding color terms in the two vocabularies.}
Pairs below a root-mean squared distance threshold ($10~\Delta\mathrm{E}^*$)\footnote{Beyond this threshold vocabularies appeared visibly distinct.} were connected, and the connected components of the resulting graph determined the set of non-unique successor vocabularies.

\subsection*{Generating precursor $n - 1$ vocabularies}\label{methods:ancestral}

Candidate ($n-1$)-word precursor vocabularies for a given $n$-word rate-distortion efficient vocabulary were generated by systematically deleting each of the $n$ words in turn, and running forward the rate-distortion dynamics (Eqs.~\ref{eq:vocab-dynm-1}--\ref{eq:vocab-dynm-3}) to equilibrium.
Unique candidates (determined with the same procedure used when identifying successor vocabularies; see Methods:~\nameref{methods:successor}) were retained, and were each used in turn to generate candidate ($n-1+1$)-word successor vocabularies.
The proportion of successor vocabularies that converge to the original $n$-word vocabulary give an estimate of the relative likelihood of each candidate precursor vocabulary.
Distance to the original $n$-word vocabulary was computed as the minimum mean squared matching distance between the two vocabularies focal colors, as above.
For a given language, let $i$ index the candidate precursor vocabularies, $w_{ij}$ denote the proportion of term additions that result in the ($n-1+1$)-word vocabulary $j$, and $V_{ij}$ the distance between the original $n$-word vocabulary and the $j$th ($n-1+1$)-word vocabulary.
We estimated the overall likelihood, $p_i$, of the precursor vocabulary as
\begin{align}
    p_i &\propto \int \pi(\sigma) \sum_j w_{ij} e^{-\sigma V_{ij}} d\sigma,\label{eq:likelihood}
\end{align}
where $\sigma$ is a scaling parameter on the $n-1+1$ to $n$ distances, and $\pi(\sigma)$ is a uniform prior on $\sigma$ over an interval covering the typical variation in likelihoods across languages (SI:~\nameref{si-sec:sensitivity}; SI~Fig.~\ref{si-fig:sensitivity}).

\subsection*{Code and data availability}
All data and code used in this work are publicly available online.
The data is available via the WCS Data Archives hosted at \path{https://www.icsi.berkeley.edu/wcs/data.html}.
The code to reproduce this work is available on Github at \path{https://github.com/crtwomey/twomey2023}.
The results in this manuscript were generated using \texttt{R~v4.2.3} and the \texttt{targets} package for reproducibility \cite{landau2021}.

} 

\bibliographystyle{refs}
{\footnotesize \bibliography{refs}}

\newpage
\appendix
\counterwithin{figure}{section}

\section{Supplementary Information}

\subsection{Rate-distortion example}\label{si-sec:rd-example}

\begin{figure*}[htb]
\centering
\begin{tikzpicture}[scale=2.5]
  \foreach \row in {0,1,2}
  {
    \foreach \col in {0,1,2}
    {
      \filldraw (\col,\row) circle (0.05);
    }
    \draw[dashed] (-0.2,\row) -- (2.2,\row);
  }
  \foreach \col in {0,1,2}
  {
    \draw[dashed] (\col,-0.2) -- (\col,2.2);
  }
  \path (2.2,-.2) node[below,opacity=0] {baseline};
\end{tikzpicture}
\scalebox{0.7}{
\input{fig/rd_example.tex}
}
\caption{%
\textbf{%
Example rate-distortion trade-off for a set of points, $X$.
}%
({\em Left}) Points $x \in X$ arranged in a $3 \times 3$ grid in two-dimensional Euclidean space.
({\em Right}) Rate and distortion combinations in the grey shaded region are infeasible for any combination of $|\widehat{X}|$ and $\beta$, for $p(x) = 1/9$ (i.e.\ uniform communicative needs).
Curves show the efficient frontier for a given choice of $|\widehat{X}|$ (color) and any choice of $\beta$ (point on curve).
For each curve, the rate (distortion) monotonically non-decreases (non-increases) with increasing $\beta$, to the limit determined by $\log_2 |\widehat{X}|$.
For $|\widehat{X}| = |X| = 9$, as $\beta \rightarrow \infty$ the mapping, $p(\hat{x}|x)$, becomes the identity function, uniquely assigning each $x$ to each $\hat{x}$.
For $|\widehat{X}| = 1$, all $x$ are assigned to the same $\hat{x}$, independent of $\beta$.
}\label{si-fig:rd-example}
\end{figure*}
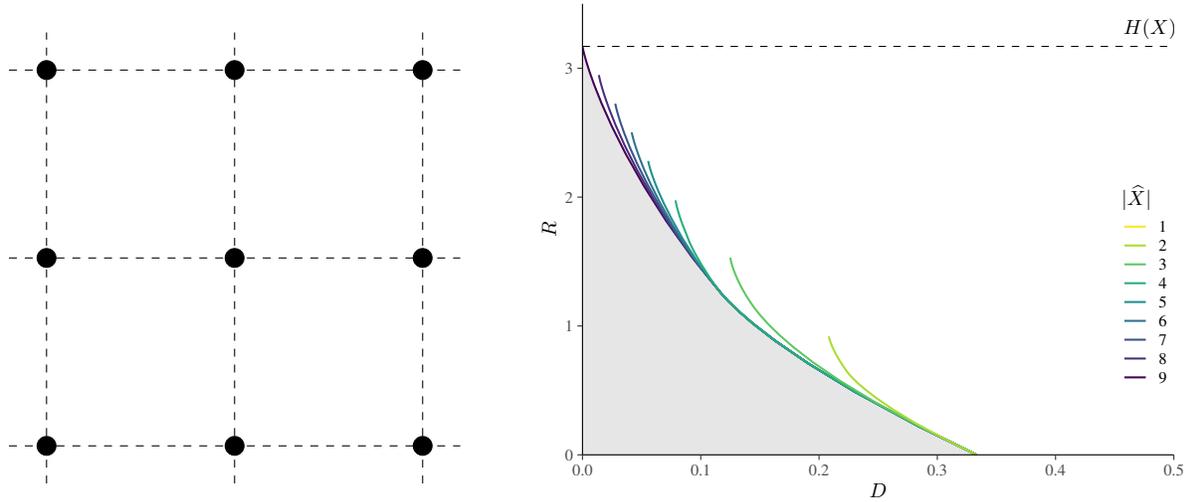

SI~Fig.~\ref{si-fig:rd-example} illustrates a simple example of a rate-distortion tradeoff for a set of points, $X$, arranged in a two-dimensional Euclidean space.
This example shows partially overlapping efficient frontiers for different choices of $\beta$ (need for precision) and vocabulary size (cardinality of the compressed representation, $\widehat{X}$).

\subsection{Sensitivity analysis}\label{si-sec:sensitivity}

\begin{figure*}[htb]
\centering
\includegraphics[width=7.5in, height=2in]{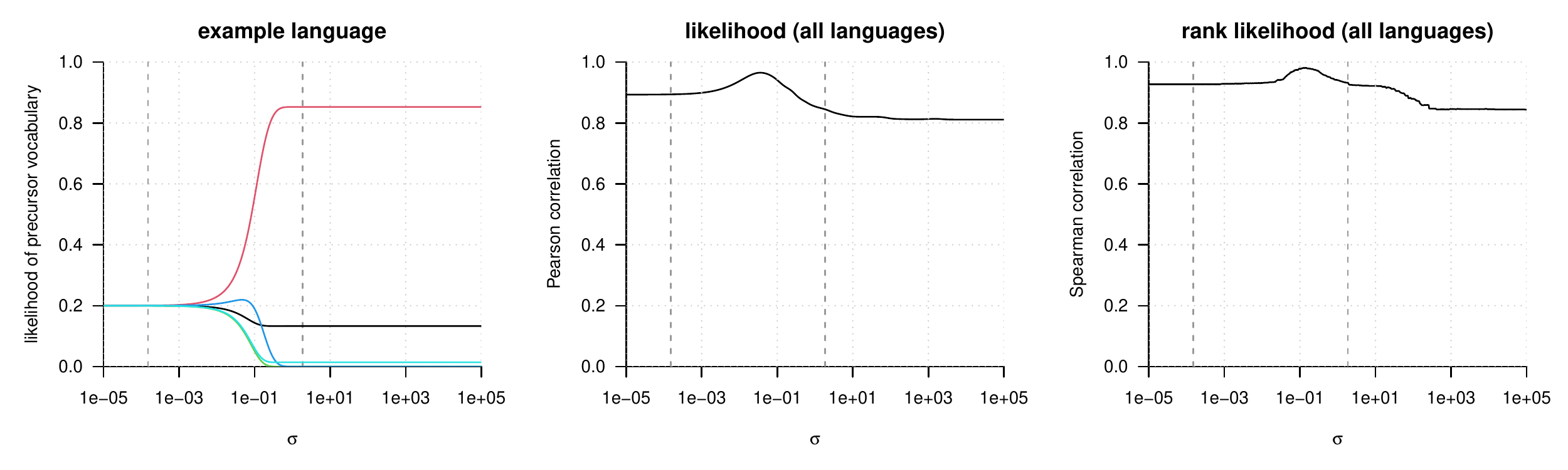}
\caption{%
\textbf{%
Sensitivity of likelihood estimates to the choice of scaling parameter, $\sigma$, in Eq.~\ref{eq:likelihood}.
}%
({\em Left}) Estimated precursor vocabulary likelihoods as a function of $\sigma$ (x-axis) for a single language.
At low values of $\sigma$, every precursor vocabulary is equally likely.
At intermediate values, precursors vary primarily by the frequency with which adding a term results in an ($n-1+1$)-word vocabulary close to the original $n$-word vocabulary.
At high values of $\sigma$, the relevance of the minimum-distance $n-1+1$ pathway is maximized.
Variation in likelihoods with respect to varying $\sigma$ primarily falls within the window demarcated by dashed vertical lines; these were the bounds used across languages for marginalizing over $\sigma$.
({\em Middle}) Pearson correlation between likelihoods computed according to Eq.~\ref{eq:likelihood} using a uniform prior over the window of variation across languages (dashed vertical lines), and a delta function at a particular choice of $\sigma$ (x-axis).
Likelihoods marginalized over the full window are broadly insensitive to non-extreme choices of $\sigma$ (i.e.\ choices that do not force likelihoods to be completely uniform or based primarily on the minimum-distance candidate).
({\em Right}) Spearman correlation between rank orderings of precursor likelihoods shows an even broader insensitivity to any particular choice of $\sigma$ (x-axis).
}\label{si-fig:sensitivity}
\end{figure*}

Precursor vocabulary likelihoods were computed according to Eq.~\ref{eq:likelihood}, which marginalizes over the choice of scale parameter, $\sigma$.
The upper and lower bounds of the uniform prior for this parameter, $\pi(\sigma)$, were chosen to span the typical window in which precursor likelihoods varied as a function of $\sigma$ (e.g.\ SI~Fig.~\ref{si-fig:sensitivity}--{\em left}).
Variation was assessed by measuring the magnitude of the change in entropy, $\h{p}$, with respect to a change in $\sigma$, i.e.\ $|\partial \h{p} / \partial \sigma|$.
The least and greatest values of $\sigma$ for which this quantity exceeded a small threshold, 1e-05, were computed for each language, and the median least and median greatest values were used as the bounds for the uniform prior over $\sigma$, used for all languages.
By marginalizing out $\sigma$ in this way, estimates of precursor likelihoods become relatively insensitive to any particular choice of $\sigma$, both in terms of the reported likelihood values (SI~Fig.~\ref{si-fig:sensitivity}--{\em middle}) and their rank order (SI~Fig.~\ref{si-fig:sensitivity}--{\em right}).

\end{document}

%% file: definitions.tex
\DeclarePairedDelimiterX{\Bregx}[2]{(}{)}{%
  #1\delimsize\|#2
}
\newcommand{\breg}{d_{\phi}\Bregx}

\DeclarePairedDelimiterX{\Dklx}[2]{(}{)}{%
  #1\delimsize\|#2
}
\newcommand{\Dkl}{D_{\mathrm{KL}}\Dklx}

\DeclarePairedDelimiterX{\Ix}[2]{(}{)}{%
  #1;#2
}
\newcommand{\I}{I\Ix}

\DeclarePairedDelimiterX{\Inx}[2]{(}{)}{%
  #1;#2
}
\newcommand{\In}{I_\mathcal{N}\Inx}

\DeclarePairedDelimiterX{\hcrossx}[2]{(}{)}{%
  #1,#2
}
\newcommand{\hcross}{H\hcrossx}

\DeclarePairedDelimiterX{\hconx}[2]{(}{)}{%
  #1|#2
}
\newcommand{\hcon}{H\hconx}

\DeclarePairedDelimiterX{\hx}[1]{(}{)}{%
  #1
}
\newcommand{\h}{H\hx}

\DeclarePairedDelimiterX{\covx}[2]{(}{)}{%
  #1,#2
}
\newcommand{\cov}{\mathrm{cov}\covx}

\DeclarePairedDelimiterX{\Itotx}[1]{(}{)}{%
  #1
}
\newcommand{\Itot}{I\Itotx}

\DeclarePairedDelimiterX{\Itotnx}[1]{(}{)}{%
  #1
}
\newcommand{\Itotn}{I_\mathcal{N}\Itotnx}

\newcommand{\stcomp}[1]{{#1}^{\mathsf{c}}}

\DeclarePairedDelimiterX{\powersetx}[1]{(}{)}{%
  #1
}
\newcommand{\powerset}{\mathcal{P}\powersetx}

\DeclarePairedDelimiterX{\setx}[1]{\{}{\}}{%
  #1
}
\newcommand{\set}{\setx}

\newcommand{\argmin}{\mathop{\mathrm{arg\,min}}}
\newcommand{\argmax}{\mathop{\mathrm{arg\,max}}}
\newcommand{\argsup}{\mathop{\mathrm{arg\,sup}}}

\DeclarePairedDelimiterX{\inner}[2]{\langle}{\rangle}{%
	#1,#2
}

\newcommand{\rand}[1]{\uppercase{#1}}
\newcommand{\mat}[1]{\mathbf{\uppercase{#1}}}
\renewcommand{\vec}[1]{\mathbf{\lowercase{#1}}}
\newcommand{\randmat}[1]{\mat{#1}}
\newcommand{\randvec}[1]{\vec{#1}}
\newcommand{\transpose}[1]{#1^\intercal}
\newcommand{\expected}[1]{\mathbb{E}#1}
\newcommand{\p}[1]{p\left(#1\right)}
\newcommand{\pest}[1]{\hat{p}\left(#1\right)}
\newcommand{\grad}{\nabla}

\DeclarePairedDelimiterX{\condpx}[2]{(}{)}{%
  #1\mid#2
}
\newcommand{\condp}{p\condpx}

\newcommand{\xhat}{\hat{x}}
\newcommand{\Xhat}{\widehat{X}}

\newcommand{\exhat}{\tilde{x}}

\DeclarePairedDelimiterX{\dphix}[2]{(}{)}{%
  #1\delimsize\|#2
}
\newcommand{\dphi}{d_\phi\dphix}

\DeclarePairedDelimiterX{\dpsix}[2]{(}{)}{%
  #1\delimsize\|#2
}
\newcommand{\dpsi}{d_\psi\dpsix}

\DeclarePairedDelimiterX{\dbregx}[2]{(}{)}{%
  #1\delimsize\|#2
}
\newcommand{\dbreg}{d\dbregx}

\newcommand{\cielab}{CIE Lab }
\newcommand{\lab}{Lab}

\newcommand{\ciedeltaE}{$\Delta\text{E}^{\star}$}

\newcommand{\Rversion}{v3.6.3}
\newcommand{\RColorscience}{\texttt{colorscience}~(v1.0.8)}
\newcommand{\Rlmefour}{\texttt{lme4}~(v1.1-21)}
\newcommand{\RDHARMa}{\texttt{DHARMa}~(v0.2.6)}
\newcommand{\Rnloptr}{\texttt{nloptr}~(v1.2.1)}
\newcommand{\Remdist}{\texttt{emdist}~(v0.3-1)}

\newcommand{\pinf}{p_\mathrm{inf}}
\newcommand{\punif}{p_\mathrm{unif}}
\newcommand{\plang}{p_\mathrm{lang}}

\makeatletter
\newcommand\addfield[3]{\expandafter\def\csname\string#1@case@#2\endcsname{#3}}
\newcommand\valuetable[2][]{%
	\newcommand#2[1]{%
		\ifcsname\string#2@case@##1\endcsname\csname\string#2@case@##1\endcsname\else#1\fi%
	}%
}
\makeatother

\theoremstyle{definition}
\newtheorem{definition}{Definition}
\newtheorem{theorem}{Theorem}
\newtheorem{lemma}{Lemma}

\renewcommand\qedsymbol{$\blacksquare$}

%% file: fig/rd_example.tex
\begin{tikzpicture}[x=1pt,y=1pt]
\definecolor{fillColor}{RGB}{255,255,255}
\path[use as bounding box,fill=fillColor,fill opacity=0.00] (0,0) rectangle (361.35,289.08);
\begin{scope}
\path[clip] (  0.00,  0.00) rectangle (361.35,289.08);
\definecolor{drawColor}{RGB}{255,255,255}
\definecolor{fillColor}{RGB}{255,255,255}

\path[draw=drawColor,line width= 0.6pt,line join=round,line cap=round,fill=fillColor] (  0.00,  0.00) rectangle (361.35,289.08);
\end{scope}
\begin{scope}
\path[clip] ( 31.81, 35.19) rectangle (351.35,279.08);
\definecolor{fillColor}{RGB}{255,255,255}

\path[fill=fillColor] ( 31.81, 35.19) rectangle (351.35,279.08);
\definecolor{fillColor}{gray}{0.90}

\path[fill=fillColor] (244.84, 35.19) --
	(244.84, 35.19) --
	(244.84, 35.19) --
	(244.84, 35.19) --
	(244.84, 35.19) --
	(244.84, 35.19) --
	(244.84, 35.19) --
	(244.84, 35.19) --
	(213.33, 50.66) --
	(178.67, 69.45) --
	(153.74, 84.66) --
	(136.13, 96.74) --
	(123.95,106.15) --
	(115.74,113.27) --
	(110.39,118.50) --
	(106.45,122.88) --
	( 91.17,141.90) --
	( 77.04,161.63) --
	( 65.02,180.52) --
	( 55.25,197.78) --
	( 47.67,212.85) --
	( 42.06,225.40) --
	( 38.10,235.35) --
	( 35.46,242.82) --
	( 33.80,248.12) --
	( 32.82,251.62) --
	( 32.28,253.78) --
	( 32.01,255.00) --
	( 31.89,255.62) --
	( 31.84,255.91) --
	( 31.82,256.02) --
	( 31.82,256.06) --
	( 31.81,256.08) --
	( 31.81,256.08) --
	( 31.81,256.08) --
	( 31.81,256.08) --
	( 31.81,256.08) --
	( 31.81,256.08) --
	( 31.81, 35.19) --
	cycle;
\definecolor{drawColor}{RGB}{68,1,84}

\path[draw=drawColor,line width= 1.1pt,line join=round] ( 31.81,256.08) --
	( 31.81,256.08) --
	( 31.81,256.08) --
	( 31.81,256.08) --
	( 31.81,256.08) --
	( 31.81,256.08) --
	( 31.82,256.06) --
	( 31.82,256.02) --
	( 31.84,255.91) --
	( 31.89,255.62) --
	( 32.01,255.00) --
	( 32.28,253.78) --
	( 32.82,251.62) --
	( 33.80,248.12) --
	( 35.46,242.82) --
	( 38.10,235.35) --
	( 42.06,225.40) --
	( 47.67,212.85) --
	( 55.25,197.78) --
	( 65.02,180.52) --
	( 77.04,161.63) --
	( 91.17,141.90) --
	(106.45,122.88) --
	(110.39,118.50) --
	(115.74,113.27) --
	(123.95,106.15) --
	(136.13, 96.74) --
	(153.74, 84.66) --
	(178.67, 69.45) --
	(213.33, 50.66) --
	(244.84, 35.19) --
	(244.84, 35.19) --
	(244.84, 35.19) --
	(244.84, 35.19) --
	(244.84, 35.19) --
	(244.84, 35.19) --
	(244.84, 35.19) --
	(244.84, 35.19);
\definecolor{drawColor}{RGB}{71,45,123}

\path[draw=drawColor,line width= 1.1pt,line join=round] ( 40.69,240.59) --
	( 40.69,240.59) --
	( 40.69,240.59) --
	( 40.69,240.59) --
	( 40.69,240.59) --
	( 40.69,240.59) --
	( 40.69,240.57) --
	( 40.70,240.51) --
	( 40.72,240.38) --
	( 40.77,240.08) --
	( 40.89,239.48) --
	( 41.14,238.35) --
	( 41.62,236.40) --
	( 42.49,233.30) --
	( 43.95,228.67) --
	( 46.25,222.16) --
	( 49.70,213.49) --
	( 54.62,202.48) --
	( 61.34,189.13) --
	( 70.11,173.63) --
	( 80.37,157.10) --
	( 92.18,140.61) --
	(106.45,122.89) --
	(110.39,118.50) --
	(115.74,113.27) --
	(123.95,106.15) --
	(136.13, 96.74) --
	(153.74, 84.66) --
	(178.67, 69.45) --
	(213.33, 50.66) --
	(244.84, 35.19) --
	(244.84, 35.19) --
	(244.84, 35.19) --
	(244.84, 35.19) --
	(244.84, 35.19);
\definecolor{drawColor}{RGB}{59,82,139}

\path[draw=drawColor,line width= 1.1pt,line join=round] ( 49.57,225.11) --
	( 49.57,225.11) --
	( 49.57,225.11) --
	( 49.57,225.11) --
	( 49.57,225.11) --
	( 49.57,225.10) --
	( 49.57,225.08) --
	( 49.58,225.02) --
	( 49.60,224.90) --
	( 49.64,224.62) --
	( 49.75,224.08) --
	( 49.97,223.08) --
	( 50.39,221.40) --
	( 51.14,218.73) --
	( 52.39,214.75) --
	( 54.36,209.15) --
	( 57.33,201.60) --
	( 61.56,192.14) --
	( 67.42,180.50) --
	( 75.20,166.74) --
	( 84.83,151.37) --
	( 95.18,136.90) --
	(106.46,122.87) --
	(110.39,118.50) --
	(115.74,113.27) --
	(123.95,106.15) --
	(136.13, 96.74) --
	(153.74, 84.66) --
	(178.67, 69.45) --
	(213.33, 50.66) --
	(244.84, 35.19) --
	(244.84, 35.19) --
	(244.84, 35.19) --
	(244.84, 35.19) --
	(244.84, 35.19) --
	(244.84, 35.19);
\definecolor{drawColor}{RGB}{44,114,142}

\path[draw=drawColor,line width= 1.1pt,line join=round] ( 58.44,209.62) --
	( 58.44,209.62) --
	( 58.44,209.62) --
	( 58.44,209.62) --
	( 58.44,209.62) --
	( 58.44,209.62) --
	( 58.44,209.60) --
	( 58.45,209.56) --
	( 58.47,209.45) --
	( 58.51,209.23) --
	( 58.59,208.77) --
	( 58.78,207.94) --
	( 59.13,206.51) --
	( 59.77,204.24) --
	( 60.82,200.86) --
	( 62.47,196.18) --
	( 64.99,189.75) --
	( 68.60,181.68) --
	( 73.66,171.62) --
	( 80.35,159.61) --
	( 88.89,146.21) --
	( 98.10,133.28) --
	(106.64,122.66) --
	(110.39,118.50) --
	(115.74,113.27) --
	(123.95,106.15) --
	(136.13, 96.74) --
	(153.74, 84.66) --
	(178.67, 69.45) --
	(213.33, 50.66) --
	(244.84, 35.19);

\path[draw=drawColor,line width= 1.1pt,line join=round] (244.84, 35.19) --
	(244.84, 35.19) --
	(244.84, 35.19) --
	(244.84, 35.19) --
	(244.84, 35.19) --
	(244.84, 35.19);
\definecolor{drawColor}{RGB}{33,144,140}

\path[draw=drawColor,line width= 1.1pt,line join=round] ( 67.32,194.14) --
	( 67.32,194.14) --
	( 67.32,194.14) --
	( 67.32,194.14) --
	( 67.32,194.14) --
	( 67.32,194.13) --
	( 67.32,194.12) --
	( 67.32,194.09) --
	( 67.34,194.01) --
	( 67.37,193.83) --
	( 67.44,193.47) --
	( 67.59,192.80) --
	( 67.87,191.64) --
	( 68.36,189.87) --
	( 69.20,187.21) --
	( 70.55,183.40) --
	( 72.63,178.17) --
	( 75.71,171.29) --
	( 80.07,162.63) --
	( 85.84,152.25) --
	( 93.32,140.51) --
	(101.18,129.47) --
	(106.83,122.43) --
	(110.39,118.50) --
	(115.74,113.27) --
	(123.95,106.15) --
	(136.13, 96.74) --
	(153.74, 84.66) --
	(178.67, 69.45) --
	(213.33, 50.66) --
	(244.84, 35.19) --
	(244.84, 35.19) --
	(244.84, 35.19) --
	(244.84, 35.19) --
	(244.84, 35.19) --
	(244.84, 35.19) --
	(244.84, 35.19);
\definecolor{drawColor}{RGB}{39,173,129}

\path[draw=drawColor,line width= 1.1pt,line join=round] ( 82.11,172.81) --
	( 82.11,172.81) --
	( 82.11,172.81) --
	( 82.11,172.80) --
	( 82.11,172.79) --
	( 82.11,172.77) --
	( 82.12,172.71) --
	( 82.14,172.58) --
	( 82.18,172.34) --
	( 82.25,171.91) --
	( 82.39,171.22) --
	( 82.62,170.15) --
	( 83.01,168.59) --
	( 83.62,166.41) --
	( 84.54,163.47) --
	( 85.88,159.69) --
	( 87.74,154.99) --
	( 89.94,149.42) --
	( 92.38,144.58) --
	( 95.88,138.40) --
	(100.79,130.69) --
	(105.03,124.67) --
	(107.04,122.19) --
	(110.39,118.50) --
	(115.74,113.27) --
	(123.95,106.15) --
	(136.13, 96.74) --
	(153.74, 84.66) --
	(178.67, 69.45) --
	(213.33, 50.66);

\path[draw=drawColor,line width= 1.1pt,line join=round] (244.84, 35.19) --
	(244.84, 35.19) --
	(244.84, 35.19) --
	(244.84, 35.19) --
	(244.84, 35.19) --
	(244.84, 35.19) --
	(244.84, 35.19) --
	(244.84, 35.19);
\definecolor{drawColor}{RGB}{93,200,99}

\path[draw=drawColor,line width= 1.1pt,line join=round] (111.70,141.83) --
	(111.70,141.82) --
	(111.70,141.81) --
	(111.70,141.77) --
	(111.71,141.70) --
	(111.73,141.57) --
	(111.76,141.37) --
	(111.81,141.07) --
	(111.90,140.62) --
	(112.04,139.98) --
	(112.26,139.11) --
	(112.58,137.94) --
	(113.07,136.41) --
	(113.76,134.43) --
	(114.75,131.96) --
	(116.07,129.01) --
	(117.77,125.63) --
	(119.93,121.80) --
	(122.72,117.43) --
	(125.03,114.19) --
	(128.13,110.35) --
	(132.36,105.69) --
	(138.08,100.10) --
	(145.80, 93.39) --
	(156.28, 85.29) --
	(170.67, 75.41) --
	(190.65, 63.22) --
	(218.66, 48.04) --
	(244.84, 35.19) --
	(244.84, 35.19) --
	(244.84, 35.19) --
	(244.84, 35.19) --
	(244.84, 35.19) --
	(244.84, 35.19) --
	(244.84, 35.19) --
	(244.84, 35.19) --
	(244.84, 35.19) --
	(244.84, 35.19);
\definecolor{drawColor}{RGB}{170,220,50}

\path[draw=drawColor,line width= 1.1pt,line join=round] (164.95, 99.18) --
	(164.95, 99.17) --
	(164.95, 99.17) --
	(164.96, 99.16) --
	(164.96, 99.13) --
	(164.97, 99.08) --
	(164.98, 98.98) --
	(165.02, 98.81) --
	(165.08, 98.53) --
	(165.18, 98.12) --
	(165.35, 97.53) --
	(165.60, 96.71) --
	(165.99, 95.61) --
	(166.57, 94.16) --
	(167.42, 92.27) --
	(168.64, 89.83) --
	(170.39, 86.74) --
	(172.83, 82.92) --
	(174.94, 79.93) --
	(175.94, 78.69) --
	(177.62, 76.84) --
	(180.29, 74.23) --
	(184.39, 70.67) --
	(190.48, 65.97) --
	(199.29, 59.92) --
	(211.75, 52.32) --
	(229.09, 42.92) --
	(244.84, 35.19) --
	(244.84, 35.19) --
	(244.84, 35.19) --
	(244.84, 35.19) --
	(244.84, 35.19) --
	(244.84, 35.19) --
	(244.84, 35.19) --
	(244.84, 35.19) --
	(244.84, 35.19) --
	(244.84, 35.19);
\definecolor{drawColor}{RGB}{253,231,37}

\path[draw=drawColor,line width= 1.1pt,line join=round] (244.84, 35.19) --
	(244.84, 35.19) --
	(244.84, 35.19) --
	(244.84, 35.19) --
	(244.84, 35.19) --
	(244.84, 35.19) --
	(244.84, 35.19) --
	(244.84, 35.19) --
	(244.84, 35.19) --
	(244.84, 35.19) --
	(244.84, 35.19) --
	(244.84, 35.19) --
	(244.84, 35.19) --
	(244.84, 35.19) --
	(244.84, 35.19) --
	(244.84, 35.19) --
	(244.84, 35.19) --
	(244.84, 35.19) --
	(244.84, 35.19) --
	(244.84, 35.19) --
	(244.84, 35.19) --
	(244.84, 35.19) --
	(244.84, 35.19) --
	(244.84, 35.19) --
	(244.84, 35.19) --
	(244.84, 35.19) --
	(244.84, 35.19) --
	(244.84, 35.19) --
	(244.84, 35.19) --
	(244.84, 35.19) --
	(244.84, 35.19) --
	(244.84, 35.19) --
	(244.84, 35.19) --
	(244.84, 35.19) --
	(244.84, 35.19) --
	(244.84, 35.19) --
	(244.84, 35.19) --
	(244.84, 35.19) --
	(244.84, 35.19) --
	(244.84, 35.19) --
	(244.84, 35.19);
\definecolor{drawColor}{RGB}{0,0,0}

\path[draw=drawColor,line width= 0.3pt,dash pattern=on 4pt off 4pt ,line join=round] ( 31.81,256.08) -- (351.35,256.08);

\node[text=drawColor,anchor=base,inner sep=0pt, outer sep=0pt, scale=  1.10] at (338.57,263.05) {$H(X)$};
\end{scope}
\begin{scope}
\path[clip] (  0.00,  0.00) rectangle (361.35,289.08);
\definecolor{drawColor}{RGB}{0,0,0}

\path[draw=drawColor,line width= 0.6pt,line join=round] ( 31.81, 35.19) --
	( 31.81,279.08);
\end{scope}
\begin{scope}
\path[clip] (  0.00,  0.00) rectangle (361.35,289.08);
\definecolor{drawColor}{gray}{0.30}

\node[text=drawColor,anchor=base east,inner sep=0pt, outer sep=0pt, scale=  0.88] at ( 26.86, 32.16) {0};

\node[text=drawColor,anchor=base east,inner sep=0pt, outer sep=0pt, scale=  0.88] at ( 26.86,101.84) {1};

\node[text=drawColor,anchor=base east,inner sep=0pt, outer sep=0pt, scale=  0.88] at ( 26.86,171.52) {2};

\node[text=drawColor,anchor=base east,inner sep=0pt, outer sep=0pt, scale=  0.88] at ( 26.86,241.21) {3};
\end{scope}
\begin{scope}
\path[clip] (  0.00,  0.00) rectangle (361.35,289.08);
\definecolor{drawColor}{gray}{0.20}

\path[draw=drawColor,line width= 0.6pt,line join=round] ( 29.06, 35.19) --
	( 31.81, 35.19);

\path[draw=drawColor,line width= 0.6pt,line join=round] ( 29.06,104.87) --
	( 31.81,104.87);

\path[draw=drawColor,line width= 0.6pt,line join=round] ( 29.06,174.55) --
	( 31.81,174.55);

\path[draw=drawColor,line width= 0.6pt,line join=round] ( 29.06,244.24) --
	( 31.81,244.24);
\end{scope}
\begin{scope}
\path[clip] (  0.00,  0.00) rectangle (361.35,289.08);
\definecolor{drawColor}{RGB}{0,0,0}

\path[draw=drawColor,line width= 0.6pt,line join=round] ( 31.81, 35.19) --
	(351.35, 35.19);
\end{scope}
\begin{scope}
\path[clip] (  0.00,  0.00) rectangle (361.35,289.08);
\definecolor{drawColor}{gray}{0.20}

\path[draw=drawColor,line width= 0.6pt,line join=round] ( 31.81, 32.44) --
	( 31.81, 35.19);

\path[draw=drawColor,line width= 0.6pt,line join=round] ( 95.72, 32.44) --
	( 95.72, 35.19);

\path[draw=drawColor,line width= 0.6pt,line join=round] (159.63, 32.44) --
	(159.63, 35.19);

\path[draw=drawColor,line width= 0.6pt,line join=round] (223.54, 32.44) --
	(223.54, 35.19);

\path[draw=drawColor,line width= 0.6pt,line join=round] (287.44, 32.44) --
	(287.44, 35.19);

\path[draw=drawColor,line width= 0.6pt,line join=round] (351.35, 32.44) --
	(351.35, 35.19);
\end{scope}
\begin{scope}
\path[clip] (  0.00,  0.00) rectangle (361.35,289.08);
\definecolor{drawColor}{gray}{0.30}

\node[text=drawColor,anchor=base,inner sep=0pt, outer sep=0pt, scale=  0.88] at ( 31.81, 24.18) {0.0};

\node[text=drawColor,anchor=base,inner sep=0pt, outer sep=0pt, scale=  0.88] at ( 95.72, 24.18) {0.1};

\node[text=drawColor,anchor=base,inner sep=0pt, outer sep=0pt, scale=  0.88] at (159.63, 24.18) {0.2};

\node[text=drawColor,anchor=base,inner sep=0pt, outer sep=0pt, scale=  0.88] at (223.54, 24.18) {0.3};

\node[text=drawColor,anchor=base,inner sep=0pt, outer sep=0pt, scale=  0.88] at (287.44, 24.18) {0.4};

\node[text=drawColor,anchor=base,inner sep=0pt, outer sep=0pt, scale=  0.88] at (351.35, 24.18) {0.5};
\end{scope}
\begin{scope}
\path[clip] (  0.00,  0.00) rectangle (361.35,289.08);
\definecolor{drawColor}{RGB}{0,0,0}

\node[text=drawColor,anchor=base,inner sep=0pt, outer sep=0pt, scale=  1.10] at (191.58, 12.14) {$D$};
\end{scope}
\begin{scope}
\path[clip] (  0.00,  0.00) rectangle (361.35,289.08);
\definecolor{drawColor}{RGB}{0,0,0}

\node[text=drawColor,rotate= 90.00,anchor=base,inner sep=0pt, outer sep=0pt, scale=  1.10] at ( 17.58,157.13) {$R$};
\end{scope}
\begin{scope}
\path[clip] (  0.00,  0.00) rectangle (361.35,289.08);
\definecolor{fillColor}{RGB}{255,255,255}

\path[fill=fillColor] (317.70, 64.35) rectangle (353.05,176.74);
\end{scope}
\begin{scope}
\path[clip] (  0.00,  0.00) rectangle (361.35,289.08);
\definecolor{drawColor}{RGB}{0,0,0}

\node[text=drawColor,anchor=base west,inner sep=0pt, outer sep=0pt, scale=  1.10] at (323.20,170.08) {$|\widehat{X}|$};
\end{scope}
\begin{scope}
\path[clip] (  0.00,  0.00) rectangle (361.35,289.08);

\path[] (323.91,152.05) rectangle (336.94,165.09);
\end{scope}
\begin{scope}
\path[clip] (  0.00,  0.00) rectangle (361.35,289.08);
\definecolor{drawColor}{RGB}{253,231,37}

\path[draw=drawColor,line width= 1.1pt,line join=round] (324.64,158.57) -- (336.21,158.57);
\end{scope}
\begin{scope}
\path[clip] (  0.00,  0.00) rectangle (361.35,289.08);

\path[] (323.91,141.87) rectangle (336.94,154.90);
\end{scope}
\begin{scope}
\path[clip] (  0.00,  0.00) rectangle (361.35,289.08);
\definecolor{drawColor}{RGB}{170,220,50}

\path[draw=drawColor,line width= 1.1pt,line join=round] (324.64,148.38) -- (336.21,148.38);
\end{scope}
\begin{scope}
\path[clip] (  0.00,  0.00) rectangle (361.35,289.08);

\path[] (323.91,131.68) rectangle (336.94,144.71);
\end{scope}
\begin{scope}
\path[clip] (  0.00,  0.00) rectangle (361.35,289.08);
\definecolor{drawColor}{RGB}{93,200,99}

\path[draw=drawColor,line width= 1.1pt,line join=round] (324.64,138.20) -- (336.21,138.20);
\end{scope}
\begin{scope}
\path[clip] (  0.00,  0.00) rectangle (361.35,289.08);

\path[] (323.91,121.50) rectangle (336.94,134.53);
\end{scope}
\begin{scope}
\path[clip] (  0.00,  0.00) rectangle (361.35,289.08);
\definecolor{drawColor}{RGB}{39,173,129}

\path[draw=drawColor,line width= 1.1pt,line join=round] (324.64,128.01) -- (336.21,128.01);
\end{scope}
\begin{scope}
\path[clip] (  0.00,  0.00) rectangle (361.35,289.08);

\path[] (323.91,111.31) rectangle (336.94,124.34);
\end{scope}
\begin{scope}
\path[clip] (  0.00,  0.00) rectangle (361.35,289.08);
\definecolor{drawColor}{RGB}{33,144,140}

\path[draw=drawColor,line width= 1.1pt,line join=round] (324.64,117.83) -- (336.21,117.83);
\end{scope}
\begin{scope}
\path[clip] (  0.00,  0.00) rectangle (361.35,289.08);

\path[] (323.91,101.12) rectangle (336.94,114.16);
\end{scope}
\begin{scope}
\path[clip] (  0.00,  0.00) rectangle (361.35,289.08);
\definecolor{drawColor}{RGB}{44,114,142}

\path[draw=drawColor,line width= 1.1pt,line join=round] (324.64,107.64) -- (336.21,107.64);
\end{scope}
\begin{scope}
\path[clip] (  0.00,  0.00) rectangle (361.35,289.08);

\path[] (323.91, 90.94) rectangle (336.94,103.97);
\end{scope}
\begin{scope}
\path[clip] (  0.00,  0.00) rectangle (361.35,289.08);
\definecolor{drawColor}{RGB}{59,82,139}

\path[draw=drawColor,line width= 1.1pt,line join=round] (324.64, 97.45) -- (336.21, 97.45);
\end{scope}
\begin{scope}
\path[clip] (  0.00,  0.00) rectangle (361.35,289.08);

\path[] (323.91, 80.75) rectangle (336.94, 93.78);
\end{scope}
\begin{scope}
\path[clip] (  0.00,  0.00) rectangle (361.35,289.08);
\definecolor{drawColor}{RGB}{71,45,123}

\path[draw=drawColor,line width= 1.1pt,line join=round] (324.64, 87.27) -- (336.21, 87.27);
\end{scope}
\begin{scope}
\path[clip] (  0.00,  0.00) rectangle (361.35,289.08);

\path[] (323.91, 70.57) rectangle (336.94, 83.60);
\end{scope}
\begin{scope}
\path[clip] (  0.00,  0.00) rectangle (361.35,289.08);
\definecolor{drawColor}{RGB}{68,1,84}

\path[draw=drawColor,line width= 1.1pt,line join=round] (324.64, 77.08) -- (336.21, 77.08);
\end{scope}
\begin{scope}
\path[clip] (  0.00,  0.00) rectangle (361.35,289.08);
\definecolor{drawColor}{RGB}{0,0,0}

\node[text=drawColor,anchor=base west,inner sep=0pt, outer sep=0pt, scale=  0.88] at (343.15,155.54) {1};
\end{scope}
\begin{scope}
\path[clip] (  0.00,  0.00) rectangle (361.35,289.08);
\definecolor{drawColor}{RGB}{0,0,0}

\node[text=drawColor,anchor=base west,inner sep=0pt, outer sep=0pt, scale=  0.88] at (343.15,145.35) {2};
\end{scope}
\begin{scope}
\path[clip] (  0.00,  0.00) rectangle (361.35,289.08);
\definecolor{drawColor}{RGB}{0,0,0}

\node[text=drawColor,anchor=base west,inner sep=0pt, outer sep=0pt, scale=  0.88] at (343.15,135.17) {3};
\end{scope}
\begin{scope}
\path[clip] (  0.00,  0.00) rectangle (361.35,289.08);
\definecolor{drawColor}{RGB}{0,0,0}

\node[text=drawColor,anchor=base west,inner sep=0pt, outer sep=0pt, scale=  0.88] at (343.15,124.98) {4};
\end{scope}
\begin{scope}
\path[clip] (  0.00,  0.00) rectangle (361.35,289.08);
\definecolor{drawColor}{RGB}{0,0,0}

\node[text=drawColor,anchor=base west,inner sep=0pt, outer sep=0pt, scale=  0.88] at (343.15,114.80) {5};
\end{scope}
\begin{scope}
\path[clip] (  0.00,  0.00) rectangle (361.35,289.08);
\definecolor{drawColor}{RGB}{0,0,0}

\node[text=drawColor,anchor=base west,inner sep=0pt, outer sep=0pt, scale=  0.88] at (343.15,104.61) {6};
\end{scope}
\begin{scope}
\path[clip] (  0.00,  0.00) rectangle (361.35,289.08);
\definecolor{drawColor}{RGB}{0,0,0}

\node[text=drawColor,anchor=base west,inner sep=0pt, outer sep=0pt, scale=  0.88] at (343.15, 94.42) {7};
\end{scope}
\begin{scope}
\path[clip] (  0.00,  0.00) rectangle (361.35,289.08);
\definecolor{drawColor}{RGB}{0,0,0}

\node[text=drawColor,anchor=base west,inner sep=0pt, outer sep=0pt, scale=  0.88] at (343.15, 84.24) {8};
\end{scope}
\begin{scope}
\path[clip] (  0.00,  0.00) rectangle (361.35,289.08);
\definecolor{drawColor}{RGB}{0,0,0}

\node[text=drawColor,anchor=base west,inner sep=0pt, outer sep=0pt, scale=  0.88] at (343.15, 74.05) {9};
\end{scope}
\end{tikzpicture}